\documentclass[10pt, conference, compsocconf,final]{IEEEtran}

\usepackage[ruled,linesnumbered]{algorithm2e}
\usepackage{hyperref}
\usepackage{amsmath,amsthm,amssymb}

\newcommand{\BO}[1]{{O}\left(#1\right)}
\newcommand{\BT}[1]{{\Theta}\left(#1\right)}
\newcommand{\BOM}[1]{{\Omega}\left(#1\right)}

\newtheorem{lemma}{Lemma}
\newtheorem{theorem}{Theorem}

\newtheorem{corollary}{Corollary}

\newcommand{\sn}{\sqrt{n}}
\newcommand{\sm}{\sqrt{m}}
\newcommand{\tn}{\tilde{n}}
\renewcommand{\to}{\tilde{o}}

\renewcommand{\d}{\delta}
\newcommand{\e}{\epsilon}

\def\MR{MR$(m,M)$}
\def\A{\mathcal{A}}

\begin{document}

\title{Space-Round Tradeoffs for MapReduce Computations}

\author{\IEEEauthorblockN{Andrea Pietracaprina\IEEEauthorrefmark{1},
Geppino
Pucci\IEEEauthorrefmark{1}, Matteo Riondato\IEEEauthorrefmark{2}, Francesco
Silvestri\IEEEauthorrefmark{1} and Eli Upfal\IEEEauthorrefmark{2}}
\IEEEauthorblockA{\IEEEauthorrefmark{1}Dipartimento di Ingegneria dell'Informazione \\
Universit\`a di Padova, \\
Padova, Italy \\
Email: \{capri,geppo,silvest1\}@dei.unipd.it\\}
\IEEEauthorblockA{\IEEEauthorrefmark{2}Department of Computer Science \\
Brown University \\
Providence, RI USA \\
Email: \{matteo,eli\}@cs.brown.edu}}

\maketitle

\begin{abstract}
This work explores fundamental modeling and algorithmic issues arising in
the well-established MapReduce framework. First, we formally specify
a computational model for MapReduce which captures the functional
flavor of the paradigm by allowing for a flexible use of parallelism.
Indeed, the model diverges from a traditional processor-centric view
by featuring parameters which embody only global and local memory
constraints, thus favoring a more data-centric view. Second, we
apply the model to the fundamental computation task of matrix multiplication presenting
upper and lower bounds for both dense
and sparse matrix multiplication, which highlight interesting
tradeoffs between space and round complexity. Finally, building on the
matrix multiplication results, we derive further space-round tradeoffs
on matrix inversion and matching.
\end{abstract}


\section{Introduction}
In recent years, MapReduce has emerged as a computational paradigm for
processing large-scale data sets in a
series of rounds executed on conglomerates of commodity servers \cite{DeanG08}, and  has
been widely adopted by a
number of large Web companies (e.g., Google, Yahoo!, Amazon) and in
several other applications (e.g., GPU and multicore processing).
(See \cite{LinD10} and references therein.)

Informally, a MapReduce computation transforms
an input set of key-value pairs into an output set of key-value pairs
in a number of \emph{rounds}, where in each round  each
pair is first individually transformed into a (possibly empty) set of new pairs
(\emph{map step}) and then all values associated with the same key are
processed, separately for each key, by an instance of the same reduce function
(simply called \emph{reducer} in the rest of the paper) thus producing the next new set of
key-value pairs
(\emph{reduce step}). In fact, as already noticed in \cite{LattanziMSV11}, a
reduce step can clearly embed the subsequent map step so that a
MapReduce computation can be simply seen as a sequence of rounds of
(augmented) reduce steps.

The MapReduce paradigm has a functional flavor, in that it merely
requires that the algorithm designer decomposes the computation into
rounds and, within each round, into independent tasks through the use
of keys. This enables parallelism without forcing an algorithm to
cater for the explicit allocation of processing
resources. Nevertheless, the paradigm implicitly posits the existence
of an underlying  unstructured and possibly heterogeneous parallel
infrastructure, where the computation is eventually run.  While mostly ignoring
the details of such an underlying infrastructure, existing formalizations
of the MapReduce paradigm constrain the computations to abide with some
local and aggregate memory limitations.

In this paper, we look at both modeling and algorithmic issues related
to the MapReduce paradigm. We first provide a formal
specification of the model, aimed at overcoming some limitations of the
previous modeling efforts, and then derive interesting tradeoffs
between memory constraints and round complexity for the fundamental
problem of matrix multiplication and some of its applications.

\subsection{Previous work}

The MapReduce paradigm has been introduced in \cite{DeanG08} without a
fully-specified formal computational model for algorithm design and
analysis. Triggered by the popularity quickly gained by the paradigm,
a number of subsequent works have dealt more rigorously with
modeling and algorithmic issues~\cite{KarloffSV10,GoodrichSZ11,FeldmanMSSS10}.

In \cite{KarloffSV10}, a MapReduce algorithm specifies a sequence of
rounds as described in the previous section. Somewhat arbitrarily, the
authors impose that in each round the memory needed by any reducer to
store and transform its input pairs has size $\BO{n^{1-\epsilon}}$,
and that the aggregate memory used by all reducers has size
$\BO{n^{2-2\epsilon}}$, where $n$ denotes the input size and
$\epsilon$ is a fixed constant in $(0,1)$. The cost of local
computation, that is, the work performed by the individual reducers,
is not explicitly accounted for, but it is required to be polynomial
in $n$. The authors also postulate, again somewhat arbitrarily, that
the underlying parallel infrastructure consists of
$\BT{n^{1-\epsilon}}$ processing elements with $\BT{n^{1-\epsilon}}$
local memory each, and hint at a possible way of supporting the
computational model on such infrastructure, where the reduce instances
are scheduled among the available machines so to distribute the
aggregate memory in a balanced fashion. It has to be remarked that
such a distribution may hide non negligible costs for very
fine-grained computations (due to the need of allocating multiple
reducer with different memory requirements to a fixed number of
machines) when, in fact, the algorithmic techniques of
\cite{KarloffSV10} do not fully explore the larger power of the
MapReduce model with respect to a model with fixed parallelism.  In
\cite{LattanziMSV11} the same model of \cite{KarloffSV10} is adopted
but when evaluating an algorithm the authors also consider the total
work and introduce the notion of work-efficiency typical of the
literature on parallel algorithms.

An alternative computational model for MapReduce is  proposed in
\cite{GoodrichSZ11},  featuring two parameters which describe
bandwidth and latency characteristics of the underlying communication
infrastructure, and an additional parameter that limits the amount of
I/O performed by each reducer.  Also, a BSP-like cost function is
provided which combines the internal work of the reducers with the
communication costs incurred by the shuffling of the data needed at
each round. Unlike the model of \cite{KarloffSV10}, no limits are posed
to the aggregate memory size. This implies that in principle there is
no limit to the allowable parallelism while, however, the
bandwidth/latency parameters must somewhat reflect the topology and,
ultimately, the number of processing elements. Thus, the model mixes
the functional flavor of MapReduce with the more descriptive nature of
bandwidth-latency models such as BSP \cite{Valiant90,BilardiP11}.

A model which tries to merge the spirit of MapReduce with the features
of data-streaming  is the MUD model of \cite{FeldmanMSSS10}, where the reducers
receive their input key-value pairs as a stream to be processed in one pass
using small working memory, namely polylogarithmic in the input size.
A similar model has been  adopted in \cite{ChierichettiKT10}.

MapReduce algorithms for a variety of problems have been developed on
the aforementioned MapReduce variants including, among others,
primitives such as prefix sums, sorting, random indexing
\cite{GoodrichSZ11}, and graph problems such as triangle counting
\cite{TsourakakisKMF09} minimum spanning tree, $s$-$t$ connectivity,
\cite{KarloffSV10}, maximal and approximate maximum matching, edge
cover, minimum cut \cite{LattanziMSV11}, and max cover
\cite{ChierichettiKT10}. Moreover simulations of the PRAM and BSP in
MapReduce have been presented in \cite{KarloffSV10,GoodrichSZ11}.  In
particular, it is shown that a $T$-step EREW PRAM algorithm can be
simulated by an $\BO{T}$-round MapReduce algorithm, where each reducer
uses constant-size memory and the aggregate memory is proportional to
the amount of shared memory required by the PRAM algorithm
\cite{KarloffSV10}.  The simulation of CREW or CRCW PRAM algorithms
incurs a further $\BO{\log_m (M/m)}$ slowdown, where $m$ denotes the
local memory size available for each reducer and $M$ the aggregate
memory size \cite{GoodrichSZ11}.

All of the aforementioned algorithmic efforts have been aimed at
achieving the minimum number of rounds, possibly constant, provided
that enough local memory for the reducer (typically, sublinear yet
polynomial in the input size) and enough aggregate memory is
available. However, so far, to the best of our knowledge, there has
been no attempt to fully explore the tradeoffs that can be exhibited
for specific computational problems between the local and aggregate
memory sizes, on one side, and the number of rounds, on the other,
under reasonable constraints of the amount of total work performed by
the algorithm. Our results contribute to filling this gap.

Matrix multiplication is a building block for many problems, including
matching~\cite{MulmuleyVV87}, matrix inversion~\cite{JaJa92},
all-pairs shortest path~\cite{JaJa92}, graph
contraction~\cite{GilbertSR08}, cycle detection~\cite{YusterZ04}, and
parsing context free languages~\cite{Penn06}. Parallel algorithms for
matrix multiplication of dense matrices have been widely studied:
among others, we remind \cite{IronyTT04,McCollT99} which provide upper
and lower bounds exposing a tradeoff between communication complexity
and processor memory. For sparse matrices, interesting results are
given in \cite{MiddendorfSST,Manzini94} for some network topologies
like hypercubes, in \cite{KruskalRS89} for PRAM, and in
\cite{BulucG08} for a BSP-like model. In particular, techniques in
\cite{McCollT99,KruskalRS89} are used in the following sections for
deriving efficient MapReduce algorithms. In the sequential settings,
some interesting works providing upper and lower bounds are \cite{HongK81,Kerr70} for
dense matrix multiplication, and \cite{Gustavson78,YusterU95,GreinerJ10}  for
sparse matrix multiplication.

\subsection{New results}
The contribution of this paper is twofold, since it targets both
modeling and algorithmic issues.

We first
formally specify a computational model
for MapReduce which captures the functional flavor of the paradigm by
allowing a flexible use of parallelism.  More specifically, our model
generalizes the one proposed in \cite{KarloffSV10} by letting the
local and aggregate memory sizes be two independent parameters, $m$
and $M$, respectively. Moreover our model makes no assumption on the
underlying execution infrastructure, for instance it does not impose a
bound on the number of available machines, thus fully decoupling the
degree of parallelism exposed by a computation from the one of the
machine where the computation will be eventually executed. This
decoupling greatly simplifies algorithm design, which has been one of
the original objectives of the MapReduce paradigm.
(In Section~\ref{sec:preliminary}, we
quantify the cost of implementing a round of our model on a
system with fixed parallelism.)

Our algorithmic contributions concern the study of attainable
tradeoffs in MapReduce for several variants of the fundamental
primitive of matrix multiplication. In particular, building on the
well-established three-dimensional algorithmic strategy for matrix
multiplication \cite{IronyTT04}, we develop upper and lower bounds for
dense-dense matrix multiplication and provide similar bounds for
deterministic and/or randomized algorithms for sparse-sparse and
sparse-dense matrix multiplication. The algorithms are parametric in
the local and aggregate memory constraints and achieve optimal or
quasi-optimal round complexity in the entire range of variability of
such parameters.  Finally, building on the matrix multiplication
results, we derive similar space-round tradeoffs for matrix inversion
and matching, which are important by-products of matrix
multiplication.

\subsection{Organization of the paper}
The rest of the paper is structured as follows. In Section~\ref{sec:preliminary} we
introduce our computational model for MapReduce and describe important
algorithmic primitives (sorting and prefix sums) that we use in our algorithms.
Section~\ref{sec:intromatrix} deals with matrix multiplication in our model,
presenting theoretical bounds to the complexity of algorithms to solve this
problem. We apply these results in Section~\ref{sec:applications} to derive
algorithms for matrix inversion and for matching in graphs.

\section{Model definition and basic primitives}\label{sec:preliminary}

Our model is defined in terms of two integral parameters $M$ and $m$,
whose meaning will be explained below, and is named \MR.  Algorithms
specified in this model will be referred to as
\emph{MR-algorithms}. An MR-algorithm specifies a sequence of
\emph{rounds}: the $r$-th round, with $r \geq 1$ transforms a multiset
$W_r$ of key-value pairs into two multisets $W_{r+1}$ and $O_r$ of
key-value pairs, where $W_{r+1}$ is the input of the next round
(empty, if $r$ is the last round), and $O_r$ is a (possibly empty)
subset of the final output. The input of the algorithm is represented
by $W_1$ while the output is represented by $\cup_{r \geq 1} O_r$,
with $\cup$ denoting the union of multisets. The universes of keys
and values may vary at each round, and we let $U_r$ denote the universe
of keys of $W_r$. The computation performed by Round $r$ is defined by
a \emph{reducer} function $\rho_r$ which is applied independently to
each multiset $W_{r,k} \subseteq W_r$ consisting of all entries in
$W_r$ with key $k \in U_r$.

Let $n$ be the input size. The two parameters $M$ and $m$ specify the
memory requirements that each round of an MR-algorithm must
satisfy. In particular, let $m_{r,k}$ denote the space needed to
compute $\rho_r(W_{r,k})$ on a RAM with $\BO{\log n}$-bit words,
including the space taken by the input (i.e., $m_{r,k} \geq
|W_{r,k}|$) and the work space, but excluding the space taken by the
output, which contributes either to $O_r$ (i.e., the final output) or
to $W_{r+1}$.  The model imposes that $m_{r,k} \in \BO{m}$, for every
$r \geq 1$ and $k \in U_r$, that $\sum_{k \in U_r} m_{r,k} \in
\BO{M}$, for every $r \geq 1$, and that $\sum_{r\geq 1} O_r=\BO{M}$.
The complexity of an MR-algorithm is
the number of rounds that it executes in the worst case, and it is
expressed as a function of the input size $n$ and of parameters $m$
and $M$. The dependency on the parameters $m$ and $M$ allows for a
finer analysis of the cost of an MR-algorithm.

As in \cite{KarloffSV10}, we require that each reducer function runs
in time polynomial in $n$. In fact, it can be easily seen
that the model defined in \cite{KarloffSV10} is equivalent to the
\MR\ model with $m \in \BO{n^{1-\epsilon }}$ and $M\in
\BO{n^{2-2\epsilon}}$, for some fixed constant $\epsilon \in (0,1)$,
except that we eliminate the additional restrictions that the number
of rounds of an algorithm be polylogarithmic in $n$ and that the
number of physical machines on which algorithms are executed are
$\BT{n^{1-\epsilon }}$, which in our opinion should not be posed at
the model level.

Compared to the model in \cite{Goodrich10}, our \MR\ model introduces
the parameter $M$ which limits the size of the aggregate memory
required at each round, whereas in \cite{Goodrich10} this size is
virtually unbounded. Moreover, the complexity analysis in \MR\ focuses
on the tradeoffs between $m$ and $M$, on one side, and the number of
rounds on the other side, while in \cite{Goodrich10} a more complex
cost function is defined which accounts for the overall message
complexity of each round, the time complexity of each reducer
computation, and the latency and bandwidth characteristics of the
executing platform.

\subsection{Sorting and prefix sum computations}

Sorting and prefix sum primitives are used in the algorithms presented in this paper. The
input to both primitives consists of a set of $n$ key-value pairs $(i,a_i)$ with $0 \leq i
< n$ and $a_i\in S$, where $S$ denotes a suitable set. For sorting, a total order is
defined over $S$ and the output is a set of $n$ key-value pairs $(i, b_i)$, where the
$b_i$'s form a permutation of the $a_i$'s and $b_{i-1}\leq b_i$ for each $0 < i < n$. For
prefix sums, a binary associative operation $\oplus$ is defined over $S$ and the output
consists of a collection of $n$ pairs $(i, b_i)$ where $b_i = a_0 \oplus \ldots \oplus
a_{i}$, for $0 \leq i < n$.

By straightforwardly adapting the results in \cite{GoodrichSZ11}
to our model we have:
\begin{theorem}\label{prefixsorting}
The sorting and prefix sum primitives for inputs of size $n$
can be implemented in \MR\ with
round complexity
\[
\BO{\log_m n},
\]
for $M = \BT{n}$.
\end{theorem}
We remark that the each reducer in the implementation of the sorting
and prefix primitives makes use of $\BT{m}$ memory words. Hence, the
same round complexity can be achieved in a more restrictive scenario
with fixed parallelism. In fact, our \MR\ model can be simulated on a
platform with $\BT{M/m}$ processing elements, each with internal
memory of size $\BT{m}$, at the additional cost of one prefix
computation per round. Therefore, $\BO{\log_m n}$ can be regarded as
an upper bound on the relative power of our model
with respect to one  with fixed parallelism.

Goodrich~\cite{Goodrich10} claims that the round complexities stated
in Theorem~\ref{prefixsorting} are optimal for any $M=\BOM{n}$ as a
consequence of the lower bound for computing the OR of $n$ bits on the
BSP model~\cite{Goodrich99}. It can be shown that the optimality
carries through to our model where the output of a reducer is not
bounded by $m$.

\section{Matrix multiplication}\label{sec:intromatrix}

Let $A$ and $B$ be two $\sn\times\sn$ matrices and let $C = A \cdot
B$. We use $a_{i,j}, b_{i,j}$ and $c_{i,j}$, with $0\leq i,j < \sn$,
to denote the entries of $A, B$ and $C$, respectively.  In this
section we present upper and lower bounds for computing the
product $C$ in \MR. The algorithms we present envision the matrices as
conceptually divided into submatrices of size $\sm\times \sm$, and we
denote these matrices with $A_{i,j}$, $B_{i,j}$ and $C_{i,j}$,
respectively, for $0\leq i,j < \sqrt{n/m}$.  Clearly,
$C_{i,j}=\sum_{h=0}^{\sqrt{n/m}-1} A_{i,h}\cdot B_{h,j}$.

All our algorithms exploit the following partition of the
$(n/m)^{3/2}$ products between submatrices (e.g., $A_{i,h}\cdot
B_{h,j}$) into $\sqrt{n/m}$ \emph{groups}: group $G_\ell$, with
$0\leq \ell < \sqrt{n/m}$, consists of products $A_{i, h}\cdot
B_{h,j}$, for every $0\leq i,j < \sqrt{n/m}$ and for $h=
(i+j+\ell)\mod \sqrt{n/m}$. Observe that each submatrix of $A$ and $B$
occurs exactly once in each group $G_\ell$.

We focus our attention on matrices whose entries belong to a semiring
$(S, \oplus, \odot)$ such that for any $a \in S$ we have $a\odot 0 =
0$, where $0$ is the identity for $\oplus$. In this setting, efficient
matrix multiplication techniques such as Strassen's cannot be
employed.  Moreover, we assume that the inner products of any row of
$A$ and of any column of $B$ with overlapping nonzero entries never
cancel to zero, which is a reasonable assumption when computing over
natural numbers or over real numbers with a finite numerical
precision.

In our algorithms, any input matrix $X$ ($X=A,B$) is provided as a set
of key-value pairs $(k_{i,j},(i,j,x_{i,j}))$ for all elements $x_{i,j}
\neq 0$. Key $k_{i,j}$ represents a progressive index, e.g., the
number of nonzero entries preceding $x_{i,j}$ in the row-major scan of
$X$.  We call a $\sn\times \sn$ matrix \textit{dense} if the number of
its nonzero entries is $\BT{n}$, and we call it \textit{sparse}
otherwise. We suppose that $M$ is sufficiently large to contain the
input and output matrices. In what follows, we present different
algorithms tailored for the multiplication of dense-dense
(Section~\ref{sec:ddmult}), sparse-sparse (Section~\ref{sec:ssmult}),
and sparse-dense matrices (Section~\ref{sec:sdmult}). We also derive
lower bounds which demonstrate that our algorithms are either optimal
or close to optimal (Section~\ref{sec:lb}), and an algorithm for
estimating the number of nonzero entries in the product of two sparse
matrices (Section~\ref{sec:evaluation}).

\subsection{Dense-Dense Matrix Multiplication}\label{sec:ddmult}

In this section we provide a simple, deterministic algorithm for
multiplying two dense matrices, which will be proved optimal in
Subsection~\ref{sec:lb}.  The algorithm is a straightforward
adaptation of the well-established three-dimensional algorithmic
strategy for matrix multiplication of \cite{McCollT99,IronyTT04},
however we describe a few details of its implementation in \MR\ since
the strategy is also at the base of algorithms for sparse
matrices. W.l.o.g. we may assume that $m\leq 2n$, since otherwise
matrix multiplication can be executed by a trivial sequential
algorithm. We consider matrices $A$ and $B$ as decomposed into
$\sm\times\sm$ submatrices and subdivide the products between
submatrices into groups as described above.

In each round, the algorithm computes all products within
$K=\min\{M/n, \sqrt{n/m}\}$ consecutive groups, namely, at round
$r\geq 1$, all multiplications in $G_\ell$ are computed, with
$(r-1)K\leq \ell < rK$. The idea is that in a round all submatrices of
$A$ and $B$ can be replicated $K$ times  and paired in such a way that each
reducer performs a distinct multiplication in $\cup_{(r-1)K\leq \ell <
  rK} G_\ell$.  Then, each reducer sums the newly computed product to
a partial sum which accumulates all of the products contributing to
the same submatrix of $C$ belonging to groups with the same index
modulo $K$ dealt with in previous rounds. At the end of the
$\sqrt{n}/(K\sqrt{m})$-th round, all submatrix products have been
computed. The final matrix $C$ is then obtained by adding together the
$K$ partial sums contributing to each entry of $C$ through a prefix
computation\footnote{The details of the key assignments needed to
  perform the necessary data redistributions among reducers are
  tedious but straightforward, and will be provided in the full
  version of this abstract.}.  We have the following result.
\begin{theorem}\label{th:upddmult}
The above \MR-algorithm multiplies two
$\sqrt{n}\times \sqrt{n}$ dense matrices in
\[
\BT{\frac{n^{3/2}}{M\sqrt{m}}+\log_m n}
\]
rounds.
\end{theorem}
\begin{IEEEproof}
The algorithm clearly complies with the memory constraints of
\MR\ since each reducer multiplies two $\sqrt{m}\times\sqrt{m}$
submatrices and the degree of replication is such that the algorithm
never exceeds the aggregate memory bound of $M$.  Also, the
$(n/m)^{3/2}$ products are computed in $n^{3/2}/{(M\sqrt{m})}$ rounds,
while the final prefix computation requires $\BO{\log_m K +1} = \BO{\log_m
  n}$ rounds
\end{IEEEproof}
We remark that the multiplication of two $\sqrt{n}\times \sqrt{n}$
dense matrices can be performed in a constant number of rounds
whenever $m = \BOM{n^{\epsilon}}$, for constant $\epsilon > 0$, and $M
\sqrt{m} = \BOM{n^{3/2}}$.

\subsection{Sparse-Sparse Matrix Multiplication}\label{sec:ssmult}
Consider two $\sn\times \sn$ sparse matrices $A$ and $B$ and denote
with $\tn < n$ the maximum number of nonzero entries in any of the two
matrices, and with $\to$ the number of nonzero entries in the product
$C=A \cdot B$. Below, we present two deterministic MR-algorithms (D1
and D2) and a randomized one (R1), each of which turns out to be more
efficient than the others for suitable ranges of parameters. We
consider only the case $m < 2\tn$, since otherwise matrix
multiplication can be executed by a trivial one-round MR-algorithm
using only one reducer. We also assume that the value $\tn$ is
provided in input. (If this were not the case, such a value could be
computed with a simple prefix computation in $\BO{\log_m n}$ rounds,
which does not affect the asymptotic complexity of our algorithms.)
However, we do not assume that $\to$ is known in advance since, unlike
$\tn$, this value cannot be easily computed. In fact, the only source
of randomization in algorithm R1 stems from the need to estimate
$\to$.

\subsubsection{Deterministic algorithm D1}\label{sec:verysparse}
This algorithm is based on the following strategy adapted from
\cite{KruskalRS89}. For $0\leq i <\sn$, let $a_i$ (resp., $b_i$) be
the number of nonzero entries in the $i$th column of $A$ (resp., $i$th
row of $B$), and let $\Gamma_i$ be the set containing all nonzero
entries in the $i$th column of $A$ and in the $i$th row of $B$. It is
easily seen that all of the $a_ib_i$ products between entries in
$\Gamma_i$ (one from $A$ and one from $B$) must be computed.  The
algorithm performs a sequence of \emph{phases} as follows. Suppose
that at the beginning of Phase~$t$, with $t \geq 0$, all products between entries in
$\Gamma_i$, for each $i\leq r-1$ and for a suitable value $r$ (initially, $r=0$), have
been computed and added to the appropriate entries of $C$. Through a prefix computation,
Phase~$t$ computes the largest $K_t$ such that $\sum_{j=r}^{r+K_t}a_jb_j \leq M$. Then,
all products between entries in $\Gamma_j$, for every $r \leq j \leq r+K_t$, are computed
using one reducer with constant memory for each such product. The products are then added
to the appropriate entries of $C$ using again a prefix computation.

\begin{theorem}
Algorithm D1 multiplies two sparse $\sqrt{n}\times \sqrt{n}$
matrices with at most $\tn$ nonzero entries each in
\[
\BO{\left\lceil \frac{\tn\min\{\tn, \sn\}}{M} \right\rceil\log_m M}
\]
rounds, on an \MR.
\end{theorem}
\begin{IEEEproof}
The correctness is trivial and the memory constraints imposed by the
model are satisfied since in each phase at most $M$ elementary
products are performed. The theorem follows by observing that the
maximum number of elementary products is $\tn\min\{\tn, \sn\}$ and
that two consecutive phases compute at least $M$ elementary products in $\BO{\log_m M}$
rounds.
\end{IEEEproof}

\subsubsection{Deterministic algorithm D2}\label{sec:dssmult}
The algorithm exploits the same three-dimensional algorithmic strategy
used in the dense-dense case and consists of a sequence of phases. In
Phase $t$, $t \geq 0$, all $\sm\times \sm$-size products within $K_t$
consecutive groups are performed in parallel, where $K_t$ is a
phase-specific value.  Observe that the computation of all products
within a group $G_\ell$ requires space $M_{\ell} \in [\tn,\tn+\to]$,
since each submatrix of $A$ and $B$ occurs only once in $G_\ell$ and
each submatrix product contributes to a distinct submatrix of $C$.
However, the value $M_{\ell}$ can be determined in $\BT{\tn}$ space
and $\BO{\log_m n}$ rounds by ``simulating'' the execution of the
products in $G_\ell$ (without producing the output values) and adding
up the numbers of nonzero entries contributed by each product to the
output matrix.  The value $K_t$ is determined as follows. Suppose
that, at the beginning of Phase $t$, groups $G_\ell$ have been processed, for each $\ell
\leq r-1$  and for a suitable value $r$ (initially, $r=0$). The
algorithm replicates the input matrices $K'_t = \min\{M/\tn, \sqrt{n/m}\}$ times.
Subsequently, through sorting and prefix computations the algorithm computes $M_\ell$ for
each $r\leq \ell < r+K'_t$ and determines the largest $K_t \leq K'_t$ such that
$\sum_{\ell=r}^{r+K_t} M_\ell \leq M$. Then, the actual products in $G_{\ell}$, for each
$r\leq \ell \leq r+ K_t$ are executed and accumulated (again using a prefix computation)
in the output matrix $C$. We have the following theorem.
\begin{theorem}
Algorithm $D2$ multiplies  two sparse
$\sqrt{n}\times \sqrt{n}$ matrices with at most $\tn$ nonzero entries each in
\[
\BO{\left\lceil \frac{(\tn+\to)
\sqrt{n}}{M\sqrt{m}}\right\rceil\log_m M}
\]
rounds on an \MR,  where $\to$ denotes the maximum number of nonzero
entries in the output matrix.
\end{theorem}
\begin{IEEEproof}
The correctness of the algorithm is trivial.  Phase $t$ requires a
constant number of sorting and prefix computations to determine $K_t$
and to add the partial contributions to the output matrix $C$.  Since
each value $M_{\ell}$ is $\BO{\tn+\to}$ and the groups are
$\sqrt{n/m}$, clearly, $K_t = \BOM{\min\{M/(\tn+\to),\sqrt{n/m}\}}$,
and the theorem follows.
\end{IEEEproof}
We remark that the value $\to$ appearing in the stated round
complexity needs not be explicitly provided in input to the algorithm.
We also observe that with respect to Algorithm D1, Algorithm D2
features a better exploitation of the local memories available to the
individual reducers, which compute $\sm\times\sm$-size products rather
than working at the granularity of the single entries.

By suitably combining Algorithms D1 and D2, we can get
the following result.
\begin{corollary}\label{D12Round}
There is a deterministic algorithm which multiplies  two sparse
$\sqrt{n}\times \sqrt{n}$ matrices with at most $\tn$ nonzero entries
each in
\[
\BO{\left\lceil
\frac{\min\{\tn^2, \tn \sqrt{n}, (\tn+\to)\sqrt{n/m}\}}
{M}\right\rceil\log_m M}
\]
rounds on an \MR, where $\to$ denotes the maximum number of nonzero
entries in the output matrix.
\end{corollary}

\subsubsection{Randomized algorithm R1}\label{sec:rssmult}
Algorithm D2 requires $\BO{\log_m M}$ rounds in each Phase $t$ for
computing the number $K_t$ of groups to be processed.  However, if
$\to$ were known, we could avoid the computation of $K_t$ and resort
to the fixed-$K$ strategy adopted in the dense-dense case, by processing
$K = M/{(\tn+\to)}$ consecutive groups per round.  This would yield
an overall $\BO{(\tn+\to) \sqrt{n}/(M\sqrt{m})+\log_m M}$
round complexity, where the $\log_m M$ additive term accounts for the
complexity of summing up, at the end, the $K$ contributions to each
entry of $C$. However, $\to$ may not be known a priori. In this case,
using the strategy described in Section~\ref{sec:evaluation} we can
compute a value $\widehat{o}$ which is a 1/2-approximation to $\to$ with
probability at least $1-1/n$. (We say that
$\widehat{o}$ $\e$-approximates $\to$ if $\vert
\to - \widehat{o} \vert < \e \to$.) Hence, in the algorithm we can plug in
$2\widehat{o}$ as an upper bound to $\to$.
By using the result of
Theorem~\ref{otilde} with $\epsilon = 1/2$ and $\delta = 1/(2n)$, we have:
\begin{theorem} \label{R1round}
Let $m=\BOM{\log^2 n}$. Algorithm R1  multiplies
two sparse $\sqrt{n}\times \sqrt{n}$ matrices
with at most $\tn$ nonzero entries in
\[
\BO{ \frac{(\tn+\to) \sqrt{n}}{M\sqrt{m}}+\log_m M}
\]
rounds on an \MR, with probability at least $1-1/n$.
\end{theorem}

By comparing the rounds complexities stated in Corollary~\ref{D12Round} and
Theorem~\ref{R1round}, it is easily seen that the randomized algorithm R1
outperforms the deterministic strategies when $m \in (\BOM{\log^2
  n},o(M^{\epsilon}))$, for any constant $\epsilon$, $\tn\geq
\sqrt{n/m}/\log_m M$, and $\to\leq \tn\min\{\tn, \sm\}\log_m M$.  For
a concrete example, R1 exhibits better performance when $\tn >
\sqrt{n}$, $\to = \BT{\tn}$, and $m$ is polylogarithmic in $M$.
Moreover, both the deterministic and randomized strategies
can achieve a constant round complexity for suitable values
of the memory parameters.

\subsubsection{Evaluation of $\to$}\label{sec:evaluation}
Observe that a $\sqrt{n}$-approximation to $\to$ derives from the
following simple argument. Let $a_i$ and $b_i$ be the number of
nonzero entries in the $i$th column of $A$ and in the $i$th row of $B$
respectively, for each $0\leq i < \sqrt{n}$. Then, $\to\leq
\sum_{i=0}^{\sqrt{n}-1} a_ib_i \leq \to \sqrt{n}$. Evaluating the sum
requires $\BO{1}$ sorting and prefix computations, hence a
$\sqrt{n}$-approximation of $\to$ can be computed in $\BO{\log_m \tn}$
rounds. However, such an approximation is too weak for our purposes
and we show below how to achieve a tighter approximation by adapting
a strategy born in the realm of streaming algorithms.

Let $\e > 0$ and $0 < \delta<1$ be two arbitrary values.  An
$\e$-approximation to $\to$ can be derived by adapting the algorithm
of~\cite{BarYossefJKST02} for counting distinct elements in a stream
$x_0 x_1\ldots$, whose entries are in the domain $[n]=\{0,\ldots,
n-1\}$. The algorithm of~\cite{BarYossefJKST02} makes use of a very
compact data structure, customarily called \emph{sketch} in the
literature, which consists of $\Delta=\BT{\log(1/\delta)}$ lists,
$L_1, L_2, \ldots, L_{\Delta}$.  For $0\leq w < \Delta$, $L_w$
contains the $t=\BT{\lceil 1/\epsilon^2\rceil}$ distinct smallest
values of the set $\{\phi_w(x_i): i\geq 0\}$, where
$\phi_w:[n]\rightarrow [n^3]$ is a hash function picked from a
pairwise independent family.  It is shown in \cite{BarYossefJKST02}
that the median of the values $tn^3/v_0, \ldots tn^3/v_{\Delta-1}$,
where $v_w$ denotes the $t$th smallest value in $L_w$, is an
$\e$-approximation to the number of distinct elements in the stream,
with probability at least $1-\d$.  In order to compute an
$\e$-approximation of $\to$ for a product $C = A \cdot B$ of $\sqrt{n}
\times \sqrt{n}$ matrices, we can modify the algorithm as follows.
Consider the stream of values in $[n]$ where each element of the
stream corresponds to a distinct product $a_{i,h}b_{h,j}\neq 0$ and
consists of the value $j+i\sqrt{n}$. Clearly, the number of distinct
elements in this stream is exactly $\to$.  (A similar approach has
been used in \cite{AmossenCP10} in the realm of sparse boolean matrix
products.)  We now show how to implement this idea on an \MR.

The MR-algorithm is based on the crucial observation that if the
stream of values defined above is partitioned into segments, the
sketch for the entire stream can be obtained by combining the sketches
computed for the individual segments. Specifically, two sketches are
combined by merging each pair of lists with the same index and
selecting the $t$ smallest values in the merged list. The
\MR-algorithm consists of a number of phases, where each
phase, except for the last one, produces set of $M/m$ sketches, while
the last phase combines the last batch of  $M/m$ sketches into the
final sketch, and outputs the approximation to $\to$.

We refer to the partition of the matrices into $\sqrt{m} \times
\sqrt{m}$ submatrices and group the products of submatrices as done
before. In Phase~$t$, with $t \geq 1$, the algorithm processes the
products in $K=\min\{M/\tn, \sqrt{n/m}\}$ consecutive groups,
assigning each pair of submatrices in one of the $K$ groups to a
distinct reducer.  A reducer receiving $A_{i,h}$ and $B_{h,j}$, each
with at least a nonzero entry, either computes a sketch for the stream
segment of the nonzero products between entries of $A_{i,h}$ and
$B_{h,j}$, if the total number of nonzero entries of $A_{i,h}$ and
$B_{h,j}$ exceeds the size of the sketch, namely $H=\BT{(1/\epsilon^2)
  \log (1/\delta)}$ words, or otherwise leaves the two submatrices
untouched (observe that in neither case the actual product of the two
submatrices is computed). In this latter case, we refer to the pair of
(very sparse) submatrices as a \emph{pseudosketch}. At this point, the
sketches produced by the previous phase (if $t > 1$), together with
the sketches and pseudosketches produced in the current phase are
randomly assigned to $M/m$ reducers. Each of these reducers can now
produce a single sketch from its assigned pseudosketches (if any) and merge
it with all other sketches that were assigned to it.  In the last
phase ($t = \sqrt{n/m}/K$) the $M/m$ sketches are combined into the
final one through a prefix computation, and the approximation to $\to$
is computed.

\begin{theorem}\label{otilde}
Let $m=\BOM{(1/\epsilon^2)\log (1/\d)\log (n/\d)}$ and let $\e>0$ and
$0<\d<1$ be arbitrary values. Then, with probability at least
$1-2\delta$, the above algorithm computes an $\e$-approximation to
$\to$ in
\[
\BO{ \frac{\tn \sqrt{n}}{M\sqrt{m}} +\log_m M}
\]
rounds, on an \MR
\end{theorem}
\begin{IEEEproof}
The correctness of the algorithm follows from the results of
\cite{BarYossefJKST02} and the above discussion. Recall that the value
computed by the algorithm is an $\e$-approximation to $\to$ with
probability $1-\d$.  As for the rounds complexity we observe that each
phase, except for the last one, requires a constant number of rounds,
while the last one involves a prefix computation thus requiring
$\BO{\log_m M}$ rounds. We only have to make sure that in each phase
the memory constraints are satisfied (with high probability). Note
also that a sketch of size $H \leq m$ is generated either in the
presence of a pair of submatrices $A_{i,h}$, $B_{h,j}$ containing at
least $H$ entries, or within one of the $M/m$ reducers. By the choice
of $K$, it is easy to see that in any case, the overall memory
occupied by the sketches is $\BO{M}$.  As for the constraint on local
memories, a simple modification of the standard balls-into-bins
argument \cite{MitzenmacherU05} and the union bound suffices to show
that with probability $1-\d$, in every phase when sketches and
pseudosketches are assigned to $M/m$ reducers, each reducer receives
in $\BO{m+(1/\e^2)\log(1/\d)\log (n/\d)}=\BO{m}$ words. The theorem
follows.  (More details will be provided in the full version of the
paper.)
\end{IEEEproof}

\subsection{Sparse-Dense matrix multiplication}\label{sec:sdmult}
Let $A$ be a sparse $\sn\times\sn$ matrix with at most $\tn$ nonzero
entries and let $B$ be a dense $\sn\times\sn$ matrix (the symmetric
case, where $A$ is dense and $B$ sparse, is equivalent). The algorithm
for dense-dense matrix multiplication does not exploit the sparsity of
$A$ and requires $\BO{n\sn/(M\sqrt{m})+\log_m n}$
rounds. Also, if we simply plug $\tn = n$ in the complexities of the
three algorithms for the sparse-sparse case (where $\tn$ represented
the maximum number of nonzero entries of $A$ or $B$) we do not achieve
a better round complexity.  However, a careful analysis of algorithm
D1 in the sparse-dense case reveals that its round complexity is
$\BO{\lceil\tn \sn/M\rceil\log_m M}$. Therefore, by
interleaving algorithm D1 and the dense-dense algorithm we have the
following corollary.
\begin{corollary}
The multiplication on \MR\ of a sparse $\sn\times\sn$ matrix with at
most $\tn$ nonzero entries and of a dense $\sn\times\sn$ matrix
requires a number of rounds which is  the minimum
between $\BO{\left\lceil\tn\sn/M\right\rceil\log_m M} $
and $ \BO{ n\sn/(M\sm)+\log_m n}$.
\end{corollary}
Observe that the above sparse-dense strategy outperforms
all previous algorithms for instance when $\tn = o(n/(\sqrt{m} \log_m M))$.

\subsection{Lower bounds}\label{sec:lb}
In this section we provide lower bounds for dense-dense and sparse-sparse matrix
multiplication. We restrict our attention to algorithms which perform all nonzero
elementary products, that is, on \textit{conventional} matrix multiplication
\cite{IronyTT04}. Although this assumption limits the class of algorithms, ruling out
Strassen-like techniques, an elaboration of a result in~\cite{Kerr70} shows that
computing all nonzero elementary products is necessary when entries of the input matrices
are from the semirings $(\mathbb{N}, +, \cdot)$ and $(\mathbb{N}\cup\{\infty\}, \min,
+)$.\footnote{The $(\mathbb{N}\cup\{\infty\}, \min, +)$ semiring, where $\infty$ is the
identity of  the $\min$ operation, is usually adopted while computing the shortest path
matrix of a graph given its connection matrix.} Indeed, we have the following lemma which
provides a lower bound on the number of products required by an algorithm multiplying any
two matrices of size  $\sn\times\sn$, containing $\tn_A$ and $\tn_B$ nonzero entries and
where zero entries have fixed positions (a similar lemma holds for
$(\mathbb{N}\cup\{\infty\}, \min, +)$).   As a consequence of the lemma, an algorithm that
multiplies any two arbitrary matrices in the semiring $(\mathbb{N}, +, \cdot)$ must
perform all nonzero products.

\begin{lemma}
Consider an algorithm $\A$ which multiples two $\sn\times\sn$ matrices $A$ and $B$
with $\tn_A$ and $\tn_B$ nonzero entries, respectively, from the semiring $(\mathbb{N},
+, \cdot)$ and where the positions of zero entries are fixed. Then, algorithm $\A$
must perform all the nonzero elementary products.
\end{lemma}
\begin{IEEEproof}
\cite{Kerr70} shows that each $c_{i,j}$ can be computed only by summing all terms
$a_{i,h}\cdot b_{h,j}$, with $0\leq h < \sn$, if the algorithm uses only semiring
operations. The proof relies on the analysis of the output for some suitable input
matrices, and makes some assumptions that force the algorithm to compute even zero
products. However, the result still holds if we allow all the zero products to
be ignored, but  some adjustments are required. In particular, the input matrices used in
\cite{Kerr70} do not work in our scenario because may contain less than $\tn_A$ and
$\tn_B$ nonzero entries, however it is easy to find inputs with the same properties
working in our case. More details will be provided in the full version.
\end{IEEEproof}

The following theorem exhibits a tradeoff in the lower bound between the amount of
local and aggregate memory and the round complexity of an algorithm performing
conventional matrix multiplication. The proof is similar to the one proposed
in~\cite{IronyTT04} for lower bounding the communication complexity of dense-dense matrix
multiplication in a BSP-like model: however, differences
arise since we focus on round complexity and our model does not assume the outdegree of a
reducer to be bounded. In the proof of the theorem we use the following lemma which was
proved using the red-blue pebbling game in~\cite{HongK81} and then restated
in~\cite{IronyTT04} as follows.

\begin{lemma}[\cite{IronyTT04}]\label{lem:nummult}
Consider the conventional matrix multiplication $C=A\cdot B$, where $A$ and $B$ are two
arbitrary matrices. A processor that uses $N_A$ elements of $A$, $N_B$ elements of
$B$, and contributes to $N_C$ elements of $C$ can compute at most $(N_A N_B N_C)^{1/2}$
multiplication terms.
\end{lemma}

\begin{theorem}\label{th:lbddmult}
Consider an \MR-algorithm $\A$ for multiplying two $\sn\times\sn$ matrices
containing at most $\tn_A$ and $\tn_B$ nonzero entries, using conventional matrix
multiplications. Let $P$ and $\to$ denote the number of nonzero elementary products and
the  number of nonzero entries in the output matrix, respectively. Then, the round
complexity of $\A$ is
\begin{equation*}
\BOM{\left\lceil \frac{P}{M\sqrt{m}}\right\rceil+\log_m \left(\frac{P}{\to}\right)}.
\end{equation*}
\end{theorem}
\begin{IEEEproof}
Let $\A$ be an $R$-round \MR-algorithm computing $C=A\cdot B$. We prove that
$R=\BOM{P/(M\sqrt{m})}$. Consider the $r$-th round, with $1\leq
r \leq R$, and let $k$ be an arbitrary key in $U_r$ and $K_r=\vert U_r\vert$. We denote
with $o_{r,k}$ the space taken by the output of $\rho_r(W_{r,k})$ which contributes either
to $O_r$ or to $W_{r+1}$, and with $m_{r,k}$ the space needed to compute $\rho_r(W_{r,k})$
including the input and working space but excluding the output. Clearly, $m_{r,k}\leq m$,
$\sum_{k\in U_r} m_{r,k} \leq M$, and $\sum_{k\in U_r} o_{r,k} \leq \to \leq M$.

Suppose $M/K_r\geq m$. By Lemma~\ref{lem:nummult}, the reducer $\rho_r$ with input
$W_{r,k}$ can compute at most $m\sqrt{o_{r,k}}$ elementary products since $N_A, N_B\leq
m$ and $N_C\leq o_{r,k}$, where $N_A$ and $N_B$ denote the entries of $A$ and $B$ used in
$\rho_r(W_{r,k})$ and $N_C$ the entries of $C$ for which contributions are computed by
$\rho_r(W_{r,k})$. Then, the number of terms computed in the $r$-th round is at most $
\sum_{k\in U_r} m\sqrt{o_{r,k}}\leq m\sqrt{M K_r} \leq M\sqrt{m}, $ since $K_r\leq
M/m$ and the summation is maximized when $o_{r,k}=M/K_r$ for each $k\in U_r$.

Suppose now that $M/K_r < m$. Partition the keys in $U_r$ into $K'_r$ sets $S_0,\ldots
S_{K'_r-1}$ such that $m\leq \sum_{k\in S_j} m_{r,k}\leq 2m$ for each $0\leq j < K'_r$
(the lower bound may be not satisfied for $j=K'_r- 1$). Clearly, $\lfloor M/2m \rfloor
\leq K'_r \leq \lceil M/m\rceil$. By Lemma~\ref{lem:nummult}, the number of elementary
products computed by all the reducers $\rho_r(W_{r,k})$ with keys in a set $S_j$
is at most $\sum_{k\in S_j} (m_{r,k} m_{r,k} o_{r,k})^{1/2}$. Since $(xyz)^{1/2} +
(x'y'z')^{1/2}\leq ((x+x')(y+y')(z+z'))^{1/2}$ for each non negative assignment of the
$x,y,z,x',y',z'$ variables and since $\sum_{k\in S_j} m_{r,k} \leq 2m$, it follows that at
most $2m \sqrt{O_{r,j}}$ elementary products can be computed using keys in $S_j$, where
$O_{r,j}= \sum_{k\in S_j} o_{r,k}$. Therefore, the number of elementary products computed
in the $r$-th round is at most $ \sum_{j=0}^{K'_r-1} 2m \sqrt{O_{r,j}}\leq 2m \sqrt{M
K'_r} \leq 2M\sqrt{2m}, $ since $K'_r\leq \lceil M/m\rceil$ and the sum is maximized when
$O_{r,j}=M/K'_r$ for each $0\leq j < K'_r$.

Therefore, in each round $\BO{M\sqrt{m}}$ nonzero elementary products can be
computed, and then $R=\BOM{\lceil P/M\sqrt{m}\rceil}$. The second term of the lower
bound  follows since  there is at least one entry of $C$ given by the sum of $P/\to$
nonzero elementary products.
\end{IEEEproof}

We now specialize the above lower bound for algorithms for generic dense-dense and
sparse-sparse matrix multiplication.
\begin{corollary}
An \MR-algorithm for multiplying any two dense $\sn\times \sn$ matrices, using
conventional matrix multiplication, requires
$$\BOM{\frac{n^{3/2}}{M\sqrt{m}}+\log_mn}$$
rounds. On the other hand, an \MR-algorithm for multiplying any
two sparse matrices with at most $\tn$ nonzero entries requires
$$\BOM{\left\lceil\frac{\tn\min\{\tn,\sn\}}{M\sqrt{m}}\right\rceil+\log_m \tn}$$ rounds.
\end{corollary}
\begin{IEEEproof}
In the dense-dense case the lower bound follows by the above Theorem~\ref{th:lbddmult}
since we have $P=n^{3/2}$ and $\to=n$  when $\tn_A=\tn_B=n$. In the
sparse-sparse case, we set $\tn_A=\tn_B=\tn$ and we observe that there exist 
assignments of the input matrices  for which $P=\tn\min\{\tn, \sn\}$, and others
where $P/\to=\BOM{\tn}$
\end{IEEEproof}

The deterministic algorithms for matrix multiplication provided in this section perform
conventional matrix multiplication, and hence the above corollary applies. Thus, the
algorithm for dense-dense matrix multiplication described in Section~\ref{sec:ddmult} is
optimal for any value of the parameters. On the other hand, the deterministic algorithm D2
for sparse-sparse matrix multiplication given in Section~\ref{sec:dssmult} is optimal as
soon as $\tn\geq \sn$, $\to=\BO{\tn}$ and $m$ is polynomial in $M$.

\section{Applications}\label{sec:applications}

Our matrix multiplications results can be used to derive efficient
algorithms for inverting a square matrix and for solving several
variants of the matching problem in a graph. The algorithms in this
section make use of division and exponentiation. To avoid the
intricacies of dealing with limited precision, we assume each memory
word is able to store any value that occurs in the computation.  A
similar assumption is made in the presentation of algorithms for the
same problems on other parallel models (see e.g. \cite{JaJa92}).

\subsection{Inverting a lower triangular matrix}\label{sec:lowertrianinv}
In this section we study the problem of inverting a lower triangular matrix $A$ of size
$\sqrt{n}\times\sqrt{n}$. We adopt the simple recursive algorithm which leverages on the
easy formula for inverting a $2\times 2$ lower triangular matrix~\cite[Sect. 8.2]{JaJa92}.
We have
\begin{equation}\label{eq:invtriang}
\left[ \begin{array}{cc} a & 0\\ b & c \end{array} \right]^{-1} = \left[ \begin{array}{cc}
a^{-1} & 0 \\ -c^{-1}ba^{-1}  & c^{-1}
\end{array} \right].
\end{equation}
For $0\leq k \leq ({1}/{2})\log({n}/{m})$ and $0\leq i,j < 2^k$, let $A^{(k)}_{i,j}$ be
the $(i,j)$ submatrix resulting from the splitting of $A$ into submatrices of size
$(\sqrt{n}/{2^k})\times({\sqrt{n}}/{2^k})$. Since Equation~\eqref{eq:invtriang} holds even
when  $a,b,c$ are matrices, we have that $\left(A_{i,i}^{(k)}\right)^{-1}$ can be
expressed as in Equation~\eqref{eq:prodmatrinv} in Figure~\ref{fig:prodmatrinv}. Note that
$A^{-1}=\left(A^{(0)}_{0,0}\right)^{-1}$.

\begin{figure*}[t]
\begin{equation}\label{eq:prodmatrinv}
\left(A^{(k)}_{i,i}\right)^{-1} = \left[\begin{array}{cc}
  \left(A^{(k+1)}_{2i,2i}\right)^{-1} & \mathbf{0} \\
 -\left(A^{(k+1)}_{2i+1,2i+1}\right)^{-1}\cdot
A^{(k+1)}_{2i+1,2i}\cdot\left(A^{(k+1)}_{2i,2i}\right)^{-1}
 & \left(A^{(k+1)}_{2i+1,2i+1}\right)^{-1} \end{array}\right], 0\leq i\leq 2^k-1.
\end{equation}
\caption{Equation~\eqref{eq:prodmatrinv} -- Expression for
$\left(A^{(k)}_{i,i}\right)^{-1}$ }
\label{fig:prodmatrinv}
\end{figure*}

The \MR-algorithm for computing the inverse of $A$ works in $(1/2)\log (n/m)$ phases. Let
$v_r=(1/2)\log (n/m) -r$  for $0\leq r< (1/2)\log (n/m)$. In the first part of Phase $0$,
the inverses of all the lower triangular submatrices $A^{\left(v_0\right)}_{i,i}$, with
$0\leq i < \sqrt{n/m}$, are computed in parallel. Since each submatrix has size
$\sqrt{m}\times\sqrt{m}$, each inverse can be computed sequentially  within a single
reducer. In the second part of Phase $0$, each product
\[
-\left(A^{\left(v_0\right)}_{2w+1,2w+1}\right)^{-1}\cdot
A^{\left(v_0\right)}_{2w+1,2w}\cdot\left(A^{\left(v_0\right)}_{2w+1,2w+1}\right)^{-1},
\]
for $0\leq w < ({1}/{2})\sqrt{n/m}$, is computed within a reducer.

In Phase $r$, with $1\leq r < (1/2) \log (n/m)$, each term
\[
-\left(A^{\left(v_r\right)}_{2w+1,2w+1}\right)^{-1}\cdot
A^{\left(v_r\right)}_{2w+1,2w}\cdot\left(A^{\left(v_r\right)}_{2w+1,2w+1}\right)^{-1},
\]
for  $ 0\leq w <2^{v_{r+1}}$, is computed in parallel by performing two matrix
multiplications using $M/2^{v_{r+1}}$ aggregate memory and local size $m$.
Therefore, at the end of Phase $({1}/{2})\log ({n}/{m})-1$ we have all the components of
$\left(A^{(0)}_{0,0}\right)^{-1}$, i.e., of $A^{-1}$.

\begin{theorem}\label{thm:trianmatrinv}
The above algorithm computes the inverse of a nonsingular lower triangular
$\sqrt{n}\times\sqrt{n}$ matrix $A$ in
\[ \BO{\frac{n^{3/2}}{M\sqrt{m}}+\frac{\log^2 n}{\log m}}\]
rounds on an \MR.
\end{theorem}

\begin{IEEEproof}
The correctness of the algorithm follows from the correctness of~\eqref{eq:prodmatrinv}
which in turns easily follows from the correctness of the formula to invert a lower
triangular $2\times2$ matrix. From the above discussion it easy to see that the memory
requirements are all satisfied.

We now analyze the round complexity of the algorithm. At Phase $r$ we have to compute
$2^{1+v_{r+1}} =({1}/{2})^{r}\sqrt{{n}/{m}}$ products between matrices of size $
\sqrt{n}/2^{v_r}\times\sqrt{n}/2^{v_r} =2^r\sqrt{m} \times 2^r\sqrt{m}. $ Each product is
computed in parallel by using $M/2^{v_{r+1}}=M 2^{r+1}\sqrt{m/n}\geq 1$ aggregate memory
and thus each Phase $r$ requires $\BO{2^{2r} \sqrt{mn}/M + \log_m (2^r m)}$ rounds by
using the algorithm described in Section~\ref{sec:ddmult}. The cost of the lower
triangular matrix inversion algorithm is then
\[
\BO{\sum_{r=0}^{(1/2)\log(n/m)-1} \left(2^{2r} \sqrt{mn}/M +  \log_m (2^r m)\right)},
\]
which gives the bound stated in the theorem.
\end{IEEEproof}

If $M\sqrt{m}$ is $\BOM{n^{3/2}}$ and $m = \BOM{n^{\epsilon}}$ for some constant
$\epsilon$, the complexity reduces to $\BO{\log n}$ rounds, which is a logarithmic factor
better than what could be obtained by simulating the PRAM algorithm.

It is also possible to compute $A^{-1}$ using the closed formula
derived by unrolling a blocked forward substitution. In general, the
closed formula contains an exponential number of terms. There are
nonetheless special cases of matrices for which a large number of
terms in the sum are zero and only a polynomial number of terms is
left. This is, for instance, the case for triangular band matrices. (Note that
the inverse of a triangular band matrix is triangular but not necessarily a
triangular band matrix.) If the width of the band is $\BO{m\log n}$, then we
have a polynomial number of terms in the formula. In this case we can do matrix
inversion in constant rounds for sufficiently large values of $m$ and $M$. A
complete discussion of this method will be presented in the full version of the
paper.

\subsection{Inverting a general matrix}\label{sec:genmatrinv}
Building on the inversion algorithm for triangular matrices presented in the previous
subsection, and on the dense-dense matrix multiplication algorithm, in this section we
develop an \MR-algorithm to invert a general $\sqrt{n}\times\sqrt{n}$ matrix $A$. Let the
trace $tr(A)$ of $A$ be defined as $\sum_{i=0}^{n-1} a_{i,i}$, where $a_{i,i}$ denotes the
entry of $A$ on the $i$-th row and $i$-th column. The algorithm is based on the following
known strategy (see e.g., \cite[Sect.~8.8]{JaJa92}).
\begin{enumerate}
  \item Compute the powers $A^2,\dots,A^{\sqrt{n}-1}$.\label{step:inv1}
  \item Compute the traces $s_k={\sum_{i=1}^{\sqrt{n}}} tr(A^k)$, for $1\le
k\le  \sqrt{n}-1$.\label{step:inv2}
   \item Compute the coefficients $c_i$ of the characteristic polynomial of $A$
    by solving a lower triangular system of $\sqrt{n}$ linear equations
    involving the traces $s_k$ (the system is shown below).\label{step:inv3}
\item Compute $ A^{-1} = -(1/c_0)\sum_{i=1}^{\sqrt{n}} c_i A^{i-1}. $\label{step:inv4}
\end{enumerate}

We now provide more details on the MR implementation of above strategy. The algorithm
requires $M=\BOM{n^{3/2}}$, which ensures that enough aggregate memory is available to
store all the $\sqrt{n}$ powers of $A$. In Step~\ref{step:inv1}, the algorithm computes
naively the powers in the form $A^{2^i}$, $1\leq i\leq \log\sqrt{n}$, by performing a
sequence of $\log \sqrt{n}$ matrix multiplications using the algorithm in
Section~\ref{sec:ddmult}. Then, each one of the remaining powers is computed using
$M/\sqrt{n}\geq n$ aggregate memory and by performing a sequence of at most $\log\sqrt{n}$
multiplications of the matrices $A^{2^i}$ obtained earlier. In Step~\ref{step:inv2}, the
$\sqrt{n}$ traces $s_k$ are computed in parallel using a prefix like computation, while
the coefficients $c_i$ of the characteristic polynomial are computed in
Step~\ref{step:inv3} by solving the following lower triangular system:
\begin{equation*}\label{eq:matrinvsystem}
\left[
\begin{array}{ccccc}
  1 & 0 & 0 & \dots & 0 \\
  s_1 & 2 & 0 & \dots & 0 \\
  s_2 & s_1 & 3 & \ddots & \vdots \\
  \vdots & \ddots & \ddots & \ddots & 0 \\
  s_{n-1} & s_{n-2} & \dots & s_1 & n
\end{array}
\right]
\left[
\begin{array}{c}
  c_{n-1} \\
  c_{n-1} \\
  c_{n-3} \\
  \vdots \\
  c_0
\end{array}
\right]
= -
\left[
\begin{array}{c}
  s_1 \\
  s_2 \\
  s_3 \\
  \vdots \\
  s_n
\end{array}
\right].
\end{equation*}
If we
denote with $L$ the matrix on the left hand side, with $C$ the vector of unknowns, and
with $S$ the vector of the traces on the right hand side, we have $C = - L^{-1}S$.  In
order to compute the coefficients in $C$ the algorithm inverts the
$\sqrt{n}\times\sqrt{n}$ lower triangular matrix $L$ as described in
Section~\ref{sec:lowertrianinv}, and computes the product between $L^{-1}$ and $S$, to
obtain $C$. Finally, Step~\ref{step:inv4} requires a prefix like computation. We have the
following theorem.

\begin{theorem}\label{thm:genmatrinv}
The above algorithm computes the inverse of any nonsingular $\sqrt{n}\times\sqrt{n}$
matrix $A$ in
\[\BO{\frac{n^2\log n}{M\sqrt{m}} + \frac{\log^2n}{\log m}}\]
rounds on \MR, with $M=\BOM{n^{3/2}}$.
\end{theorem}

\begin{IEEEproof}
For the correctness of the algorithm see \cite[Sect.~8.8]{JaJa92}). It is easy to check
that the memory requirements of the \MR\ model are satisfied. We focus here on analyzing
the round complexity.
  
Computing the powers in the form $A^{2^i}$, $1\leq i\leq \log\sqrt{n}$ requires
$\BO{n^{3/2}\log n/(M\sqrt{m})+(\log^2 n)/ \log m}$ rounds, since the algorithm performs
a sequence of $\log \sqrt{n}$ products. The remaining
powers are computed in $\BO{n^2\log n/(M\sqrt{m})+ (\log^2 n)/(\log m)}$ rounds since
each power is computed by performing at most $\log \sqrt{n}$ product using $M/\sqrt{n}$
aggregate memory.  The prefix like computation for finding the $\sqrt{n}$ traces $s_k$
requires $\BO{\log_m n}$ rounds, while the linear system takes
$\BO{n^{3/2}/(M\sqrt{m}) + (\log^2 n)/(\log m)}$ rounds.
The final step takes $\BO{\log_m n}$ rounds using a prefix like computation. The round
complexity in the statement follows.
\end{IEEEproof}

If $M\sqrt{m}$ is $\BOM{n^2\log n}$ and $m = \BOM{n^{\epsilon}}$ for some constant
$\epsilon$, the complexity reduces to $\BO{\log n}$ rounds, which is a quadratic
logarithmic factor better than what could be obtained by simulating the PRAM algorithm.

\subsection{Approximating the inverse of a matrix}
The above algorithm for computing the inverse of any nonegative matrix requires
$M=\BOM{n^{3/2}}$. In this section we provide an \MR-algorithm providing a strong
approximation of $A^{-1}$ assuming $M=\BOM{n}$.
A matrix $B$ is a \emph{strong approximation} of the inverse of an
$\sqrt{n}\times\sqrt{n}$ matrix $A$ if $\|B-A^{-1}\|/\|A^{-1}\|\leq 2^{-n^c}$, for some
constant $c>0$. The norm $\| A\|$ of a matrix $A$ is defined as
\[
\| A \| = \max_{\mathbf{x}\neq 0} \|A \mathbf{x}\|_2/\|\mathbf{x}\|_2
\]
where $\|\cdot\|_2$ denotes the Euclidean norm of a vector.
The condition number $\kappa(A)$ of a matrix $A$ is defined as $\kappa(A) = \|
A\|\|A^{-1}\|$.

An iterative method to compute a strong approximation of the inverse of a
$\sqrt{n}\times\sqrt{n}$ matrix $A$ is proposed in~\cite[Sect.~8.8.2]{JaJa92}.
The method works as follows.
Let $B_0$ be a $\sqrt{n}\times\sqrt{n}$ matrix
satisfying the condition $\|I_{\sqrt{n}}-B_0A\|=q$ for some $0<q<1$ and where
$I_{\sqrt{n}}$ is the $\sqrt{n}\times\sqrt{n}$ identity matrix. For a
$\sqrt{n}\times\sqrt{n}$ matrix $C$ let $r(C)=I_{\sqrt{n}}-CA$. We
define $B_k=(I_{\sqrt{n}}+r(B_{k-1}))B_{k-1}$, for $k>0$. We have
\[
\frac{\|B_k-A^{-1}\|}{\|A^{-1}\|}\leq q^{2^k}.
\]
By setting $B_0 = \alpha A^T$ where $\alpha = \max_i\{\sum_{j=0}^{\sqrt{n}-1}\vert
a_{i,j}\vert\} \max_j\{\sum_{i=0}^{\sqrt{n}-1}\vert a_{i,j}\vert\}$, we have
$q=1-1/(\kappa(A)^2 n)$~\cite{PanR85}. Then, if $\kappa(A)=\BO{n^c}$ for some constant
$c\geq 0$, $B_k$ provides a strong approximation when $k=\BT{ \log n}$. From the above
discussion, it is easy to derive an efficient \MR-algorithm to compute a strong
approximation of the inverse of a matrix using the algorithm for dense matrix
multiplication in Section~\ref{sec:ddmult}. 

\begin{theorem}
The above algorithm provides a strong approximation of the inverse of any
nonegative $\sqrt{n} \times \sqrt{n}$ matrix $A$ in
\[
\BO{\frac{n^{3/2}\log n}{M\sqrt{m}}+\frac{\log^2 n}{\log m}}
\]
rounds on an \MR\ when  $\kappa(A)=\BO{n^c}$ for some constant $c\geq 0$.
\end{theorem}

\begin{IEEEproof}
The correctness of the algorithm derives from~\cite{JaJa92}. Once again we only focus on
the round complexity of the algorithm. Computing $\alpha$ requires a a constant
number of prefix like computations, and hence takes $\BO{\log_m n}$
rounds. To compute  $B_k$, $k>0$ from $B_{k-1}$, we
need the value $r(B_{k-1})$ which involves a multiplication between two
$\sqrt{n}\times\sqrt{n}$ matrices and a subtraction between two matrices. Hence, each
phase requires $\BO{n^{3/2}/(M\sqrt{m}) + \log_m n}$ rounds. Since the algorithm
terminates when $k=\BT{\log n}$, the theorem follows.
\end{IEEEproof}

\subsection{Matching of general graphs}
A strategy for computing, with
probability at least 1/2, a perfect matching of a general graph using
matrix inversion is presented in~\cite{MulmuleyVV87}. The strategy is the following:
\begin{enumerate}
  \item Let the input of the algorithm be the adjacency matrix $A$ of a graph
    $G=(V,E)$ with $\sqrt{n}$ vertices and $k$ edges. \label{step:match1}
  \item Let $B$ be the matrix obtained from $A$ by substituting the entries
    $a_{i,j}=a_{j,i}=1$ corresponding to edges in the graph with the integers
    $2^{w_{i,j}}$ and $-2^{w_{i,j}}$ respectively, for $0\leq i<j<\sqrt{n}$, where
    $w_{i,j}$ is an integer chosen independently and uniformly at random from
    $[1,2k]$. We denote the entry on the $i$th row and $j$th column of $B$ as
$b_{i,j}$.\label{step:match2}
  \item Compute the determinant $det(B)$ of $B$ and the greatest integer $w$ such that
$2^w$ divides $det(B)$.\label{step:match3}
  \item Compute $adj(B)$, the adjugate matrix of $B$, and denote the entry
  on the $i$th row and $j$th column as $adj(B)_{i,j}$.\label{step:match4}
  \item For each edge $(v_i,v_j)\in E$, compute \[ a_{i,j} = \frac{b_{i,j} \cdot
    adj(B)_{i,j}}{2^w}.\] If $a_{i,j}$ is odd, then add the edge $(v_i,v_j)$ to
    the matching.\label{step:match5}
\end{enumerate}
An \MR-algorithm for perfect matching easily follows by the above strategy. We now
provide more details on the MR implementation which assumes $M=\BOM{n^{3/2}}$.

In Step~\ref{step:match2}, $B$ is obtained as follows. The algorithm partitions $A$ into
square
$\sqrt{m}\times\sqrt{m}$
submatrices $A_{\ell, h}$, $0\leq \ell,h< \sqrt{n/m}$, and then assigns each pair of
submatrices $(A_{\ell, h },A_{h, \ell})$ to a different reducer. This assignment ensures
that each pair of entries $(a_{i,j},a_{j,i})$ of $A$ is sent to the same reducer. Consider
now the reducer receiving the pair of submatrices $(A_{\ell, h},A_{h, \ell})$ and consider
the set of pairs $(a_{i,j},a_{j,i})$ of $A$ such that $a_{i,j}=a_{j,i}=1$, where
$\ell\sqrt{m}\leq i < (\ell+1)\sqrt{m}$, $h\sqrt{m}\leq j < (h+1)\sqrt{m}$, and $i<j$. For
each of these pairs the reducer chooses a $w_{i,j}$ independently and uniformly at random
from $[1,2k]$, and sets $b_{i,j}$ to $2^{w_{i,j}}$ and $b_{j,i}$ to $-2^{w_{i,j}}$. For
all
the other entries $a_{i,j}=a_{j,i}=0$, the reducer sets $b_{i,j}=b_{j,i}=0$.

Let $c_k$, $0\leq k\le\sqrt{n}$ be the coefficients of the characteristic polynomial of
$B$, which can be computed as described in Section~\ref{sec:genmatrinv}.
Steps~\ref{step:match3} and~\ref{step:match4} can
be easily implemented since the determinant of $B$ is $c_0$ and $ adj(B) = -
(c_1I+c_2B+c_3B^2+\dots+c_{\sqrt{n}}B^{\sqrt{n}-1}).$

Finally, in  Step~\ref{step:match5}, matrices $B$ and $adj(B)$ are partitioned in square
submatrices of size $\sqrt{m}\times\sqrt{m}$, and corresponding submatrices assigned to
the same reducer, which computes the values $a_{i,j}$ for the entries in its submatrices
and outputs the edges belonging to the matching.

\begin{theorem}
The above algorithm computes, with probability at least 1/2, a perfect matching of the
vertices of a graph $G$, in
\[\BO{\frac{n^2\log n}{M\sqrt{m}} + \frac{\log^2n}{\log m}}\]
rounds on \MR, where $M=\BOM{n^{3/2}}$.
\end{theorem}
\begin{IEEEproof}
The correctness of the algorithm follows from the correctness of~\cite{MulmuleyVV87} and
it is easy to see that the memory requirements of the \MR\ model are satisfied. We focus
here on the round complexity. From the above description, it is easy to see that the
computation of $B$ and the $w_{i,j}$'s in Step~\ref{step:match2} only takes one round.
Steps~\ref{step:match3} and~\ref{step:match4} require
the computation of the coefficients of the characteristic polynomial of $B$, and so takes
a number of rounds equal to the algorithm for matrix inversion described in
Section~\ref{sec:genmatrinv}, i.e., $\BO{(n^2\log n)/(M\sqrt{m} )+ (\log^2n)/(\log m)}$.
Step~\ref{step:match5} takes one round. Since the round complexity is dominated by the
number of rounds
needed to compute the coefficients of the characteristic polynomial of $B$, the theorem
follows.
\end{IEEEproof}

We note that matching is as easy as matrix inversion In the \MR\ model. The above
result can be extend to minimum weight perfect matching, to maximum
matching, and to other variants of matching in the same way as
in~\cite[Sect.~5]{MulmuleyVV87}.

\section{Conclusions}
In this paper, we provided a formal computational model for the
MapReduce paradigm which is parametric in the local and aggregate
memory sizes and retains the functional flavor originally intended for
the paradigm, since it does not require algorithms to explicitly
specify a processor allocation for the reduce instances. Performance
in the model is represented by the round complexity, which is
consistent with the idea that when processing large data sets the
dominant cost is the reshuffling of the data. The two memory
parameters featured by the model allow the algorithm designer to
explore a wide spectrum of tradeoffs between round complexity and
memory availability.  In the paper, we covered interesting
such tradeoffs for the fundamental problem of matrix multiplication and
some of its applications. The study of similar tradeoffs for
other important applications (e.g., graph problems) constitutes
an interesting open problem.

\section*{Acknowledgments}
\addcontentsline{toc}{section}{Acknowledgment}
The work of Pietracaprina, Pucci and Silvestri was supported, in part, by MIUR of Italy
under project AlgoDEEP, and by the University of Padova under the Strategic Project
STPD08JA32 and Project CPDA099949/09. The work of Riondato and Upfal was supported, in
part, by NSF award IIS-0905553 and by the University of Padova through the Visiting
Scientist 2010/2011 grant.

\IEEEtriggeratref{21}
\bibliographystyle{IEEEtran}
% \bibliography{newalias,biblio}


\end{document}